\documentclass[aps,showpacs,preprintnumbers,amsmath, amssymb]{revtex4}

\usepackage{float}
\usepackage{graphics,epsfig}
\usepackage{graphicx}
\usepackage{dcolumn}
\usepackage{bm}
\usepackage{slashed}

\begin{document}
\baselineskip=0.8 cm

\title{{\bf Observing various phase transitions in the holographic model of superfluidity }}
\author{Xiao-Mei Kuang, Yunqi Liu,  Bin Wang}
\affiliation{INPAC, Department of Physics and
Shanghai Key Lab for Particle Physics and
Cosmology, Shanghai Jiao Tong University,
Shanghai 200240, China}

\vspace*{0.2cm}
\begin{abstract}
\baselineskip=0.6 cm
\begin{center}
{\bf Abstract}
\end{center}
   We study the gravity duals of supercurrent
solutions in the AdS black hole background with
general phase structure to describe both the
first and the second order phase transitions at
finite temperature in strongly interacting
systems. We argue that the conductivity and the
pair susceptibility can be possible
phenomenological indications to distinguish the
order of phase transitions. We  extend our
discussion to the AdS soliton configuration.
Different from the black hole spacetime, in the
probe limit the first order phase transition
cannot be brought by introducing the spatial
component of the vector potential of the gauge
field in the AdS soliton background.

\end{abstract}

\maketitle
\newpage
\vspace*{0.2cm}

The AdS/CFT
correspondence\cite{Maldacena,S.S.Gubser-1,E.Witten} has
been used to model strongly interacting systems
in terms of a gravity dual. Recently it was found
that this correspondence can provide some
insights into superconductivity \cite{S.A.
Hartnoll,C.P. Herzog,G.T. Horowitz-1}. It was
observed that a gravitational system closely
mimics the behavior of a superconductor. When the
temperature of an AdS black hole drops below the
critical value, the bulk configuration becomes
unstable and experiences a second order phase
transition. The bulk spacetime changes from
normal state to superconducting state with scalar
hair condenses on the black hole background. In
the boundary dual CFT, this corresponds to the
formation of a charged condensation. The gravity
models with the properties of holographic
superconductors have attracted considerable
interest for their potential applications to the
condensed matter physics, see for examples
\cite{G.T. Horowitz-2}-\cite{Bobev}.

In a general class of gravity duals to
superconducting theories, it was exhibited that
there exists fairly wide class of phase
transitions. It was disclosed that a generalized
St$\ddot{u}$ckelberg mechanism of symmetry
breaking allows for a description of the first
order phase transition besides the second order
phase transition \cite{S.
Franco, Q.Y. Pan-1}. Recently, in the investigation of a DC
supercurrent type solution \cite{C. Herzog-2,P.
Basu,J. Sonner,Daniel}, it was found that  the second order
superfluid phase transition can change to the
first order when the velocity of the superfluid
component increases relative to the normal
component. The novel phase diagram brought by the
supercurrent is interesting. It further enriches
the phase structure observed in the
St$\ddot{u}$ckelberg mechanism.

Whether there is an effective phenomenological
way to describe and distinguish various types of
phase transitions is a question in front of us.
In this work, we will disclose the
phenomenological signatures on various phase
transitions in the holographic model of
superfluidity. We will propose two possible probes to
distinguish the order of phase transition from
phenomenology, including the conductivity and the
pair susceptibility. We will argue that these two
quantities, which are measurable in condensed
matter physics, can help us understand more of
the phase structure in the holographic model of
superfluidity. We will present our discussions in the backgrounds of the AdS black hole and AdS soliton.

In order to have a scalar condensate in the
boundary theory, the Lagrangian with a $U(1)$
gauge field and a conformally coupled to a
charged complex scalar field $\Psi$ is expressed
in the form \cite{S.A. Hartnoll-1}
\begin{equation}\label{lag}
L=\int d^4x\sqrt{-g}\Big[-\frac{1}{4}F^{\mu \nu}F_{\mu
\nu}-|\nabla \Psi-ieA \Psi|^2-m^2|\Psi|^2\Big].
\end{equation}
To consider the possibility of DC supercurrent, both a time component $A_t$ and a spatial component $A_x$ for the vector potential have been chosen
\begin{equation}\label{vp1}
A_{\mu}=A_t(r)dt+A_x(r)dx.
\end{equation}
We are interested in static solutions and assume
all the fields are homogeneous in the field
theory direction with only radial dependence.

We will first concentrate our attention on the
four dimensional AdS black hole background with
the configuration
\begin{equation}\label{a}
ds^2=-f(r)dt^2+\frac{dr^2}{f(r)}+r^2(dx^2+dy^2)
\end{equation}
where $
f(r)=\frac{r^2}{L^2}-\frac{M}{r}
$, $L$ and $M$ are the AdS radius and the mass of the black hole.
The Hawking temperature of this black hole is read
$T=\frac{3 M^{1/3}}{4\pi L^{4/3}}$.
For the convenience of our discussion, we will set $L=1$
and the horizon $r_h=(ML^2)^{\frac{1}{3}}=1$. We will make coordinate transformation
$z=1/r$ so that the metric becomes
\begin{equation}\label{e}
ds^2=-f(z)dt^2+\frac{dz^2}{z^4f(z)}+z^{-2}(dx^2+dy^2)
\end{equation}
where $ f(z)=1/z^2-z. $ The horizon now is at
$z=1$ and the conformal boundary lies at $z=0$.

Neglecting the backreactions of the matter fields
onto the background, we have equations of motions
for fields in the probe limit
\begin{eqnarray}\label{eom1}
A_t^{''}-\frac{2\psi^2}{fz^4}A_t=0,\nonumber\\
A_x^{''}+(\frac{2}{z}+\frac{f'}{f})A_x^{'}-\frac{2\psi^2}{z^4f}A_x=0,\nonumber\\
\psi^{''}+\frac{f'}{f}\psi^{'}+\Big[\frac{(eA_t)^2}{f^2z^4}-\frac{(eA_x)^2}{fz^2}-\frac{m^2}{fz^4}\Big]\psi=0,
\end{eqnarray}
where the prime denotes the derivative with respect to $z$.

At the horizon $z=1$, the regularity requires $A_t=0$ and we have the constraints
\begin{eqnarray}\label{bdy1ps}
A_t&=&0,\nonumber\\
A_x^{'}&=&-\frac{2 \psi^2}{3z^2}A_x,\nonumber\\
\psi^{'}&=& \frac{2\psi}{3z}-\frac{1}{3}z A_x^2 \psi^2.
\end{eqnarray}
Near the AdS boundary $z\rightarrow0$, the fields behave
\begin{eqnarray}\label{bdy2}
A_t&=&\mu-\rho z+\mathcal{O}(z),\nonumber\\
A_x&=&S_x+J_x z+\mathcal{O}(z),\nonumber\\
\psi&=& z^{\vartriangle_-}\psi_1+ z^{\vartriangle_+}\psi_2+\mathcal{O}(z),
\end{eqnarray}
where $\vartriangle_\pm=\frac{3}{2}\pm
\frac{1}{2}\sqrt{9+4 m^2}$. According to the
AdS/CFT dictionary, the constant coefficients
$\mu$ and $\rho$ are the chemical potential and
the density of the charge in the dual field
theory. $J_x$ corresponds to the current and
$S_x$ gives the dual current source. Either
$\psi_1$ or $\psi_2$ is normalizable which can be
the source with the other  as the response for
the dual operator
$\psi_i\sim\mathbb{O}_i(i=1,2)$. We will
concentrate on  $\psi_1=0$ in our discussion and
set $e=1$ and $m^2=-2$ unless otherwise noted.

In the case with $A_x=0$, there appears the
second order phase transition when $1/\mu$
reaches the critical value $0.246$. Below this
critical value, the condensate starts to form.
With nonzero $A_x$, in Fig.1 we have reproduced
the result of the condensation disclosed in
\cite{C. Herzog-2,P. Basu}. For big enough
$1/\mu$, there is no condensation. When this
parameter is small enough, we observe that the
condensate does not drop to zero continuously,
this marks the first order phase transition from
the normal state to the superconducting state
when $S_x$ reaches a critical value.  For values
above the special range for $1/\mu$, the
condensation continuously drop to zero and the
phase transition between the normal state and the
superconducting state changes to the second
order. In the left panel of Fig.1 we show the
first order phase transition with $1/\mu\sim
0.146$ and in the right panel of Fig.1 we show
the second order phase transition behavior with
$1/\mu\sim 0.217$. The critical value of $1/\mu$
for the condensate to happen for $A_x\neq 0$ is
below the value for the case with $A_x=0$.
\begin{figure}\center{
\includegraphics[scale=0.8]{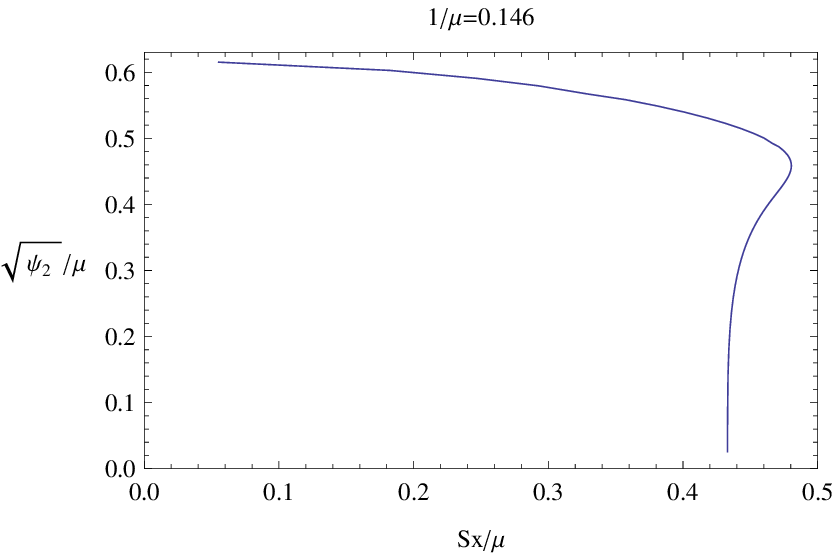}
\includegraphics[scale=0.8]{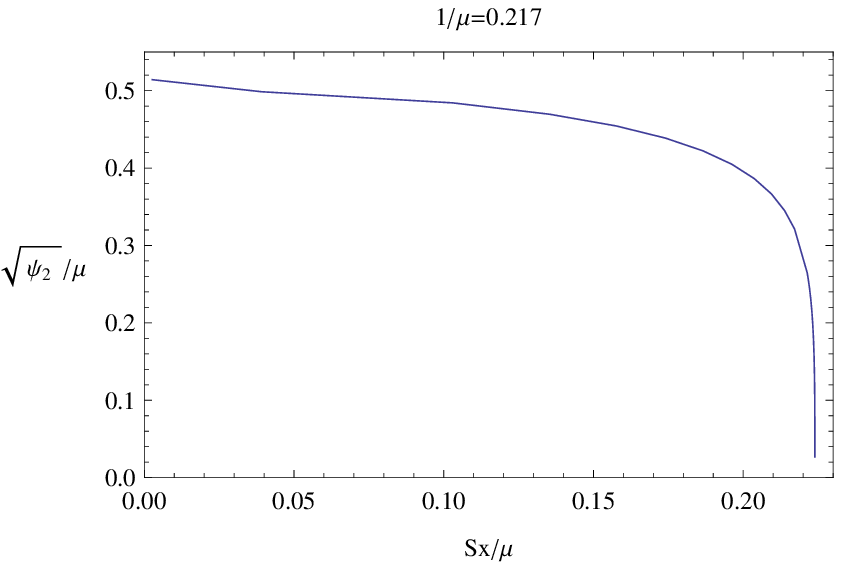}
\caption{\label{fig1} The phase structures in the
case of nonzero $A_x$. }}
\end{figure}

In the left panel of Fig.1, we observe that there
is a metastable region typical in the first order
phase transitions. This metable region also
appeared in the St$\ddot{u}$ckelberg mechanism
for the first order phase transition \cite{S.
Franco, Q.Y. Pan-1}. In this region, the
scalar field has different behavior in the
condensation. $\psi_2$ at the horizon decreases
with the decrease of $S_x/\mu$, instead of
increasing when $S_x/\mu$ becomes smaller as in
the normal condensates. In the upper panel of Fig.2 we delimit this
difference. When we use the alternative quantization, i.e. by setting $\psi_2=0$, the difference also holds in $\psi_1$
as shown in the lower panel of Fig.2.

\begin{figure} \center{
\includegraphics[scale=0.65]{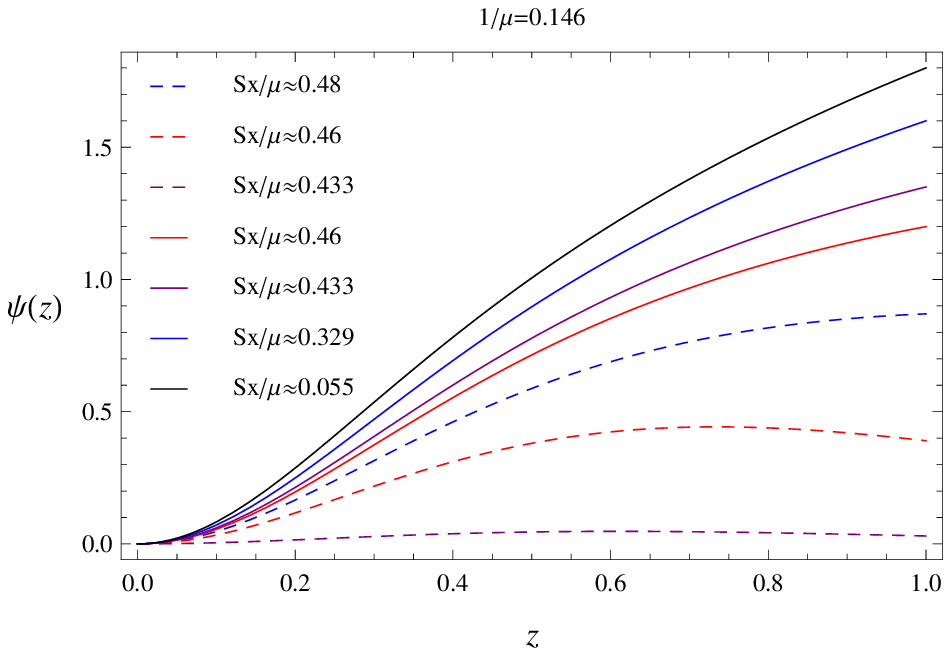}
\includegraphics[scale=0.65]{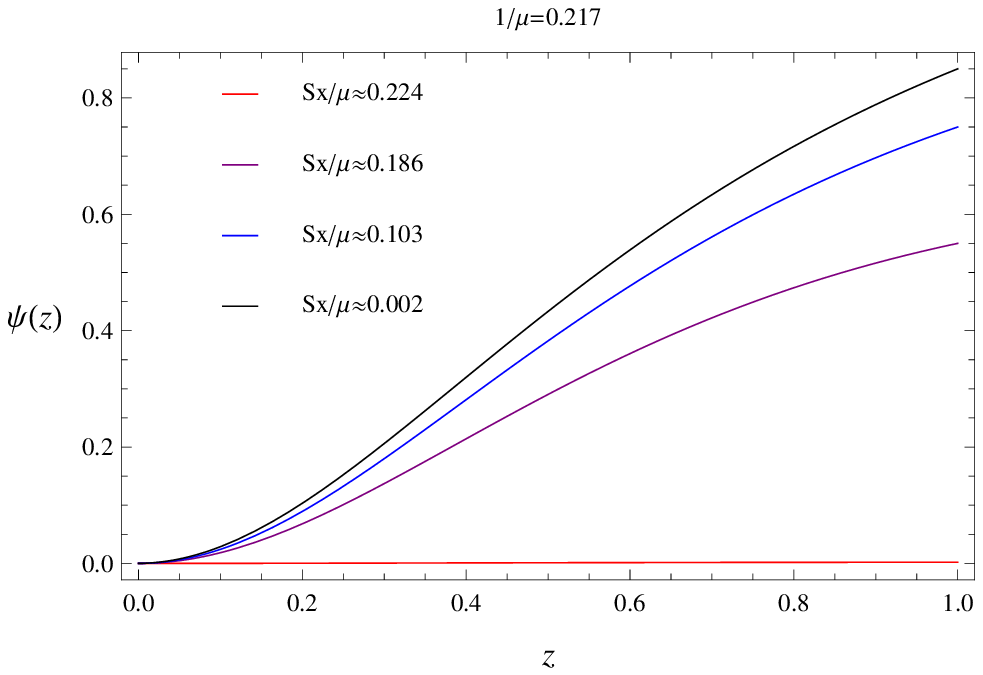}
\includegraphics[scale=0.65]{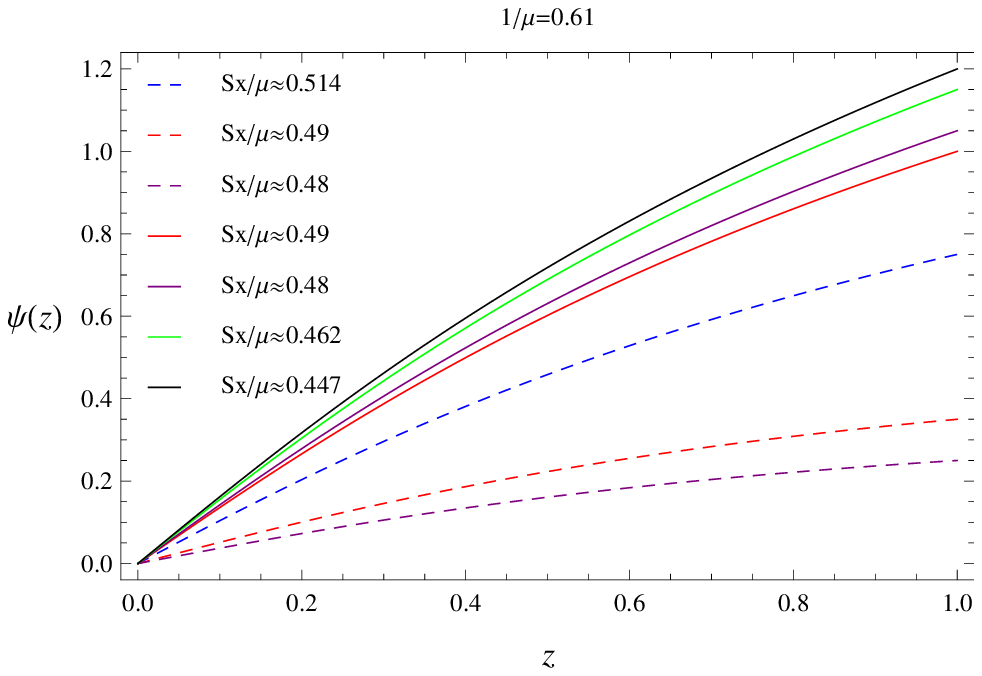}
\includegraphics[scale=0.65]{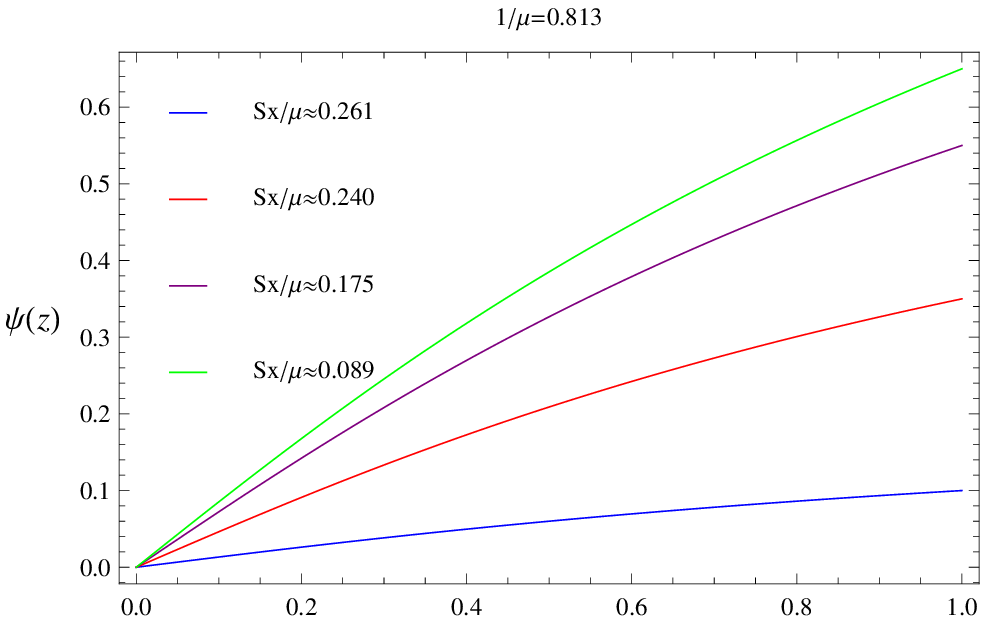}
\caption{\label{fig3}The evolution of scalar
field. The solid lines are for the stable hairy
phases while the dashed lines correspond to the
metastable states. The two figures in the upper panel
are for taking $\psi_1=0$ while those in the lower panel are for setting $\psi_2=0$.  }}
\end{figure}

In the following we will discuss the aspects of
the conductivity and examine the behaviors of
conductivity for the first order and the second
order phase transitions. We will only concentrate
on the transverse conductivity here for
simplicity and solve the transverse gauge field
perturbation $\delta A_y=e^{-iwt}A_y$ numerically
in the background with the condensate. The
equation of motion for $A_y$ is
\begin{eqnarray}\label{conduceom}
A_y^{''}+(\frac{2}{z}+\frac{f'}{f})A_y^{'}+(\frac{\omega^2}{f^2 z^4}-\frac{2\psi^2}{z^4f})A_y=0.
\end{eqnarray}
This equation can be solved by imposing the
ingoing boundary condition at the horizon for
causality. At the boundary
$A_y=A_y^{(0)}+A_y^{(1)} z+\cdots$. According to
the linear response theory, the conductivity can
be defined in terms of the retarded
current-current correlators. The conductivity can
be expressed as
\begin{eqnarray}\label{conducreal}
\sigma_y(\omega)=\frac{A_y^{(1)}}{i\omega A_y^{(0)}}.
\end{eqnarray}

In Fig.3, we plot $Re[\sigma_y]$ for the operator
$\mathbb{O}_2$. The solid lines correspond to the
second order phase transition while the dashed
lines are for the first order phase transition.
Similar to that observed in the five dimensional
situation \cite{P. Basu-1}, in our four
dimensional case we observe that there is a
conductivity gap for the first order phase
transition while the gap disappears for the
second order phase transition for the same $S_x/S_{xci}(i=1,2)$ both with $m^2=-2$ and $m^2=0$.
We can understand the nonzero conductivity gap for the first order phase
transition from two aspects. One is from the perturbation equation (\ref{conduceom}), we  see that
the profile of the scalar field imprints on the conductivity. Different profile of the field caused by different temperature and
superfluid velocity shown in Fig.2  leads to different property in conductivity.
The other is from the boundary side, the nonzero gap for the first order phase may attribute to
the release or absorption of latent heat which accompany the first order transition\cite{Binder}. 
The existence of
the conductivity gap for the first order phase
transition means that the condensate has a
binding energy, however for the second order
phase transition the condensate can have
arbitrarily low binding energy\cite{P. Basu-1}.

Furthermore we observe that the
coherence peak for the first order phase
transition is higher than that for the second
order phase transition. The difference in the coherence peak  is clearer for taking $m^2=0$.
Since the coherence peak is controlled by the thermal fluctuations of the
condensate \cite{Horbach}, this actually shows that for the
first order phase transition, the fluctuations
are stronger. In other words, higher coherence peak for the first order phase transition
indicates that only strong enough thermal fluctuations can  induce the first order
transition. Thus the superfluid velocity
and the mass of the scalar field
controls the strength of fluctuations in the
system and the strength of the coherence peak can
provide an effective description of different
orders of phase transitions.

\begin{figure} \center{
\includegraphics[scale=0.68]{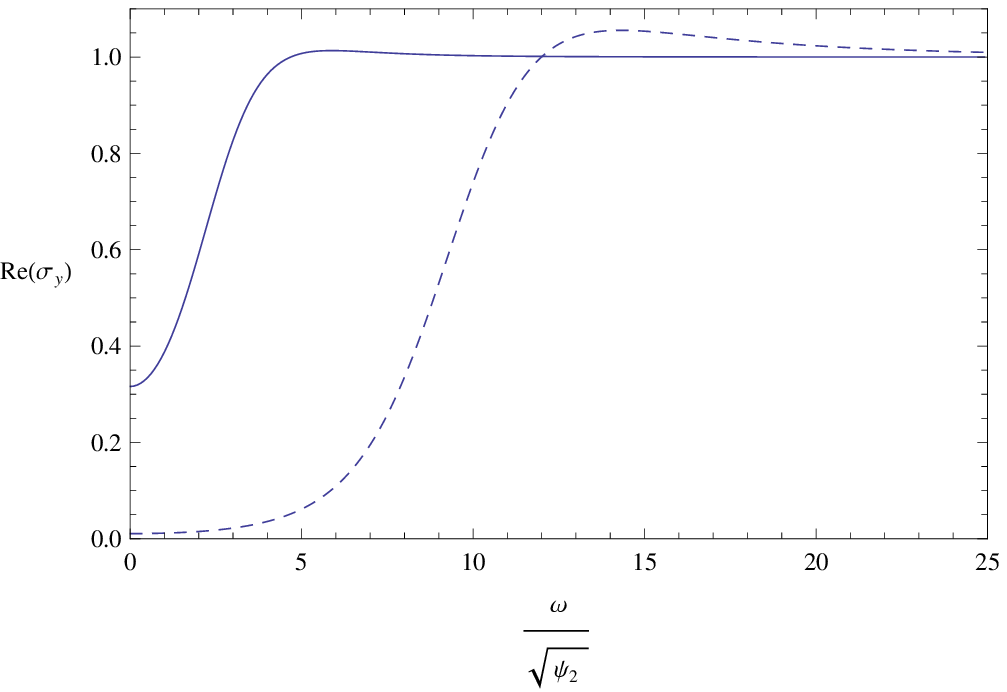}
\includegraphics[scale=0.7]{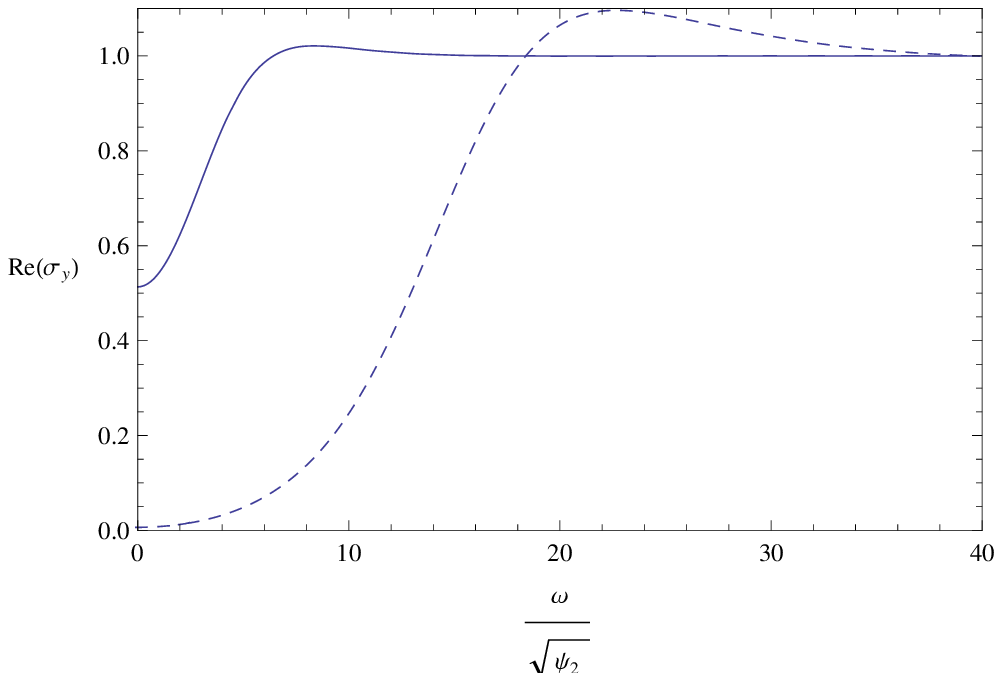}
\caption{\label{conduc} The transverse
conductivity as a function of $\omega$ for $m^2=-2$(left) and $m^2=0$(right). The
dashed lines are for the first order transition
. The solid lines are for the second order transition.}}
\end{figure}

Now we turn to discuss the susceptibility. In
\cite{K. Maeda,YQ Liu-1}, it was found that the
susceptibility can be an effective tool to probe
the holographic superconductivity. In condensed
matter physics, the dynamical pair susceptibility
can be measured directly via the second order
Josephson effect and it is believed that this
quantity can  give direct view on the origin of
the superconductivity \cite{She}. We expect that
the susceptibility can be a clear probe to
identify the order of the phase transition in the
holographic condensation.

In the dictionary of the AdS/CFT correspondence,
the dynamical susceptibility in the boundary
field theory can be calculated from the dynamics
of the fluctuations of the corresponding scalar
field in the bulk AdS background in the gravity
side. We can expand the scalar perturbation as $
\Psi=\psi(z) e^{-i \omega t}$. The equation of
motion for the scalar field reads
\begin{equation}\label{eom3}
\psi^{''}(z)+\frac{f'}{f}\psi^{'}(z)+\Big[\frac{(\omega+A_t)^2}{f^2z^4}-\frac{(A_x)^2}{fz^2}+\frac{2}{fz^4}\Big]\psi(z)=0.
\end{equation}
Note that when $\omega=0$, the equation of motion
(\ref{eom3}) goes back to the third equation of (\ref{eom1}).
Eq.(\ref{eom3}) can be solved by imposing the
infalling boundary condition at the horizon
\begin{equation}\label{hh1}
 \psi\simeq (1-z)^{(- i\omega/4\pi T)}[1+\psi^{(1)} (1-z)+\psi^{(2)} (1-z)^2+\cdots].
\end{equation}
Near the AdS boundary, the behavior of $\psi$ is
still $ \psi= z \psi_1+ z^{2} \psi_2+\cdots. $ We
choose $\psi_1$ as the source and $\psi_2$ as the
response, then the dynamical pair susceptibility
can be obtained as \cite{Son,KW}
\begin{equation}\label{chi1}
\chi=G^{R}\sim\frac{\psi_{2}}{\psi_{1}}.
\end{equation}

In the condensed matter physics,  the imaginary
part of this quantity can be measured via second
order Josephson effect and is proportional to the
current through a tunneling junction \cite{She}.

The numerical results of the imaginary part of
the dynamical pair susceptibility calculated in
our gravity background  as a function of the
frequency $\omega$ are shown in Fig.4.  We
observe that when the fluid velocity decreases
and approaches the transition point from the
normal state to the superconducting state, the
peak of the $\chi''$ becomes narrower and
stronger. Comparing the imaginary part of the
dynamical pair susceptibility exhibited in
Fig.4, we find that for
the first order phase transition the peaks of $\chi''$ are narrower than those of the second order phase transition for the same fluid velocity deviations from the critical moment. The peaks of $\chi''$ grow
more violently for the first order phase transition, which approximate five times of the corresponding peaks in the second order phase transition for the same fluid velocity deviation from the critical value. These properties also hold when we
look at the real part of the dynamical pair
susceptibility.

\begin{figure}
\center{
\includegraphics[scale=0.7]{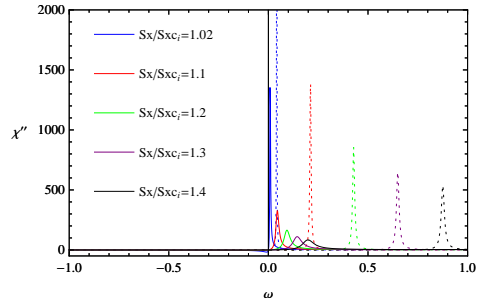}
\caption{\label{fig6}The imaginary part of the
dynamical pair susceptibility. The solid lines are for the
second order phase transition by taking
$\frac{1}{\mu}=0.217$. The dashed lines are for the
first order phase transition by taking
$\frac{1}{\mu}=0.146$. The ratio of the fluid velocity to its critical value from right to
left are $1.4$, $1.3$, $1.2$, $1.1$ and $1.02$  for the second and first order transitions.}}
\end{figure}

In addition, the static pair susceptibility
$\chi$ for $\omega=0$ can be obtained via
Kramers-Kronig relation
\begin{equation}\label{kk}
\chi|_{\omega=0}=\frac{1}{\pi} \mathcal{P}
\int_{-\infty}^{+\infty}\frac{\chi''(\omega')}{\omega'}d\omega'.
\end{equation}
 It was argued that the static
susceptibility can be an effective way to reflect
the critical behavior near the
condensation\cite{YQ Liu-1}. In the  gravity
side, to study $\chi|_{\omega=0}$, we can
numerically solve (\ref{eom3}) by setting
$\omega=0$. In Fig.5 we plot the inverse of the
static susceptibility with the change of the
ratio of the fluid velocity to its critical
value. The squares are for the first order phase
transition with $1/\mu=0.146$, while the dots are
for the second order phase transition with
$1/\mu=0.217$. It is clear that the slope for the
inverse static pair susceptibility becomes
steeper near the first order phase transition
than that near the second order phase transition.
This also confirms that the first order phase
transition has more violent phenomenon than the
second order phase transition.

Thus we find that in studying the holographic
superconductor in the AdS black hole background,
both the conductivity and  the pair
susceptibility can be helpful to distinguish the
order of the phase transition.

\begin{figure} \center{
\includegraphics[scale=0.8]{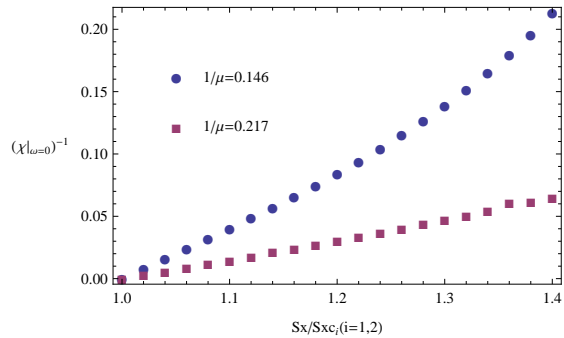}
\caption{\label{fig7}The behaviors of the static
pair susceptibility for the first order phase
transition and the second order phase
transition. We set $m^2=-2$ in the computation.}}
\end{figure}

In the following we extend our discussion to the
AdS soliton background. The soliton spacetime is
described by
\begin{equation}
ds^2=\frac{ dr^2}{ f(r)}+r^2(-dt^2+dx^2)+f(r)d\eta^2
\end{equation}
with $f(r)=r^2-\frac{r_0^3}{r}$. There is no
horizon but a tip at $r_0$. We assume the matter
fields are described by the lagrangian (\ref{lag}),
where the vector field and the scalar field are
in the form
$A_{\mu}=A_t(r)dt+A_{\eta}(r)d\eta,\space \space
\Psi=\Psi(r)$  \cite{Y. Brihaye}. The investigation
on other ansatz on the field of matters also can be
seen in \cite{Montull1,Montull2}.

In the probe limit, the equations of motions of
the matter fields in the AdS soliton background
are described by
\begin{eqnarray}\label{eom2s}
\psi^{''}+(\frac{f'}{f}+\frac{2}{r})\psi^{'}+\Big[\frac{A_t^2}{f r^2}-\frac{A_{\eta}^2}{f^2}-\frac{m^2}{f}\Big]\psi=0, \nonumber\\
A_t^{''}+\frac{f'}{f}A_t^{'}-\frac{2 \psi^2}{f}A_t=0, \nonumber\\
A_{\eta}^{''}+\frac{2}{r}A_{\eta}^{'}-\frac{2
\psi^2}{f}A_{\eta}=0,
\end{eqnarray}
where  the prime  denotes the derivative with
respect to $r$.

At the tip $r=r_0$, the fields have the
asymptotic behavior:
\begin{eqnarray}\label{eom3s}
\psi&=&U+V(r-r_0)+W(r-r_0)^2\cdots, \nonumber\\
A_t&=& M+N(r-r_0)+P(r-r_0)^2\cdots, \nonumber\\
A_{\eta}&=& a(r-r_0)+b(r-r_0)^2\cdots.
\end{eqnarray}
While near the AdS boundary the fields behave
similarly to that in the black hole background,
\begin{eqnarray}\label{eom4s}
\psi&=&\frac{\psi_1}{r}+\frac{\psi_2}{r^2}, \nonumber\\
A_t&=& \mu-\rho/r, \nonumber\\
A_{\eta}&=& S_{\eta}-J_{\eta}/r.
\end{eqnarray}
In order to compare the results with those in the
black hole, in our numerical computation we still
set $m^2=-2$ and $r_0=1$.

In Fig.6 we plot the condensation of the operator
$\mathbb{O}_2$ in the AdS Soliton background for
different values of $1/\mu$. It is different from
that we observed in the AdS black hole
background, the condensates drop to zero
continuously at critical values of the fluid
velocity. There is always second order phase
transition in the AdS soliton background when the
normal state changes to the superconducting
state. In the AdS black hole case, we know that
the first order phase transition was brought by
introducing the spatial dependence of the vector
potential and the first order structure to the
superconducting state appears at low temperature
as the fluid velocity is increased. However, in
the AdS soliton case, the spatial dependence of
the vector potential $A_{\eta}$ is not countable
to accommodate the first order phase transition,
because it behaves like $A_t$ in the AdS black
hole case. This can be easily seen by changing
$r=1/z$ in the third equation of (\ref{eom2s}).
In the AdS soliton, the first order phase
transition can not exist in the probe limit, it
can be brought only when we take account of the
strong backreaction \cite{Gary T} or in the
St$\ddot{u}$ckelberg mechanism [42]. When we
increase the values of $1/\mu$ from $1/8$, $1/5$
to $1/3$, the critical values of $S_{\eta}/\mu$
to start the condensation also increase from
$1.241$, $1.279$ to $2.641$. This is a special
character in the AdS soliton.

\begin{figure} \center{
\includegraphics[scale=0.55]{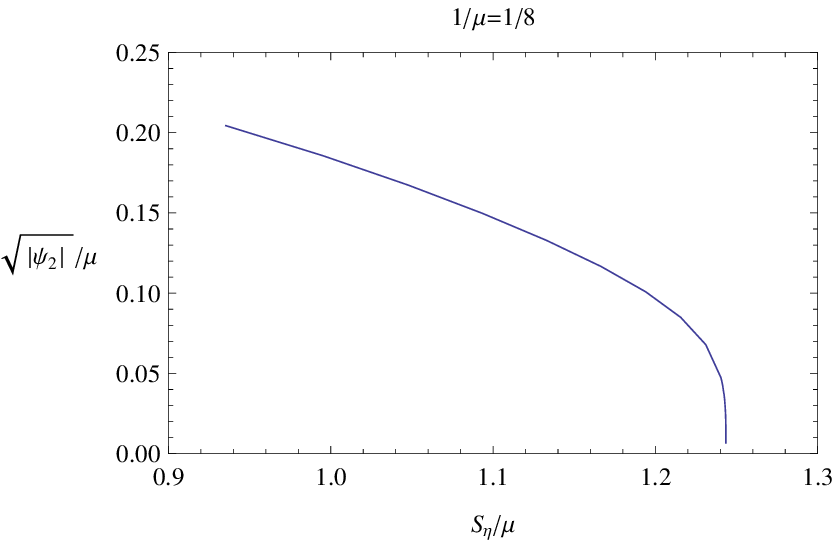}
\includegraphics[scale=0.55]{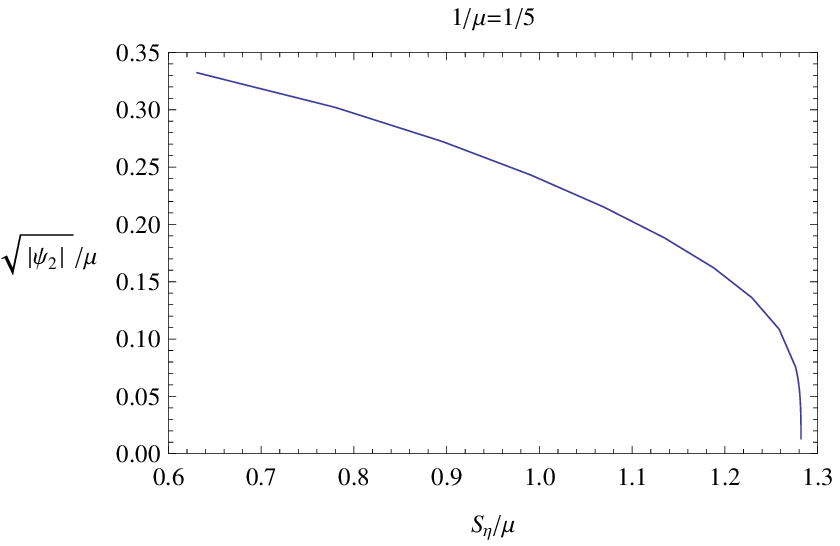}
\includegraphics[scale=0.55]{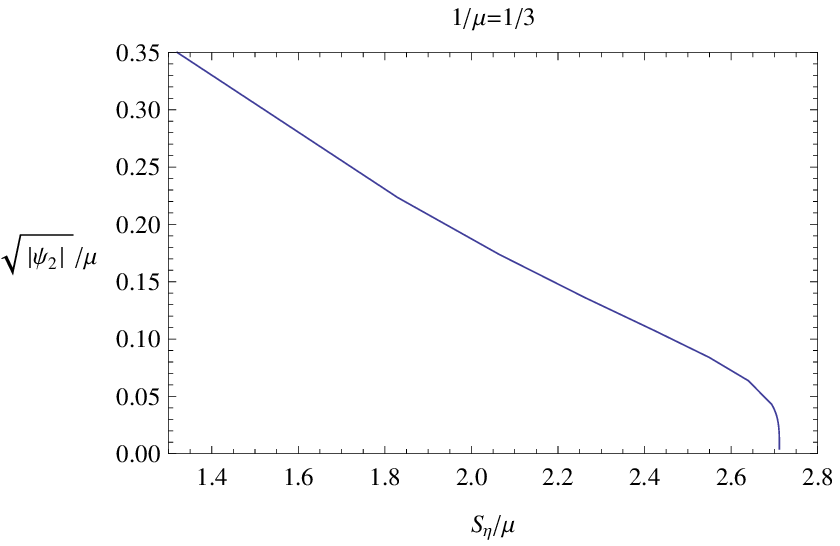}
\caption{\label{fig01} The condensates of the
scalar operator $\mathbb{O}_2$ with respect to
$S_{\eta}/\mu$ for chosen values of $1/\mu$.}}
\end{figure}

We can calculate the conductivity $\sigma_x$ when
we have the spatial dependence of the vector
potential in the AdS soliton background.
Considering the electromagnetic perturbation
$\delta A_x=e^{-i\omega t}A_x$ in the bulk, we
have the equation of motion for $A_x$
\begin{equation}\label{condsoliton}
A_x^{''}+\frac{f'}{f}A_x^{'}+(\frac{\omega^2}{fr^2}-\frac{2 \psi^2}{f})A_x=0.
\end{equation}
At the tip, $A_x$ behaves as
\begin{equation}
A_x=aa+bb(r-r_0)+cc(r-r_0)^2\cdots.
\end{equation}
While near the boundary
\begin{equation}
A_x=A_x^{(0)}+\frac{A_x^{(1)}}{r}+\cdots.
\end{equation}
The behavior of the conductivity
$\sigma_x(\omega)=\frac{A_x^{(1)}}{i\omega
A_x^{(0)}}$ is shown in Fig.7. The left panel is
for the AdS soliton before condensation while the
right panel is after the condensation.
\begin{figure} \center{
\includegraphics[scale=0.7]{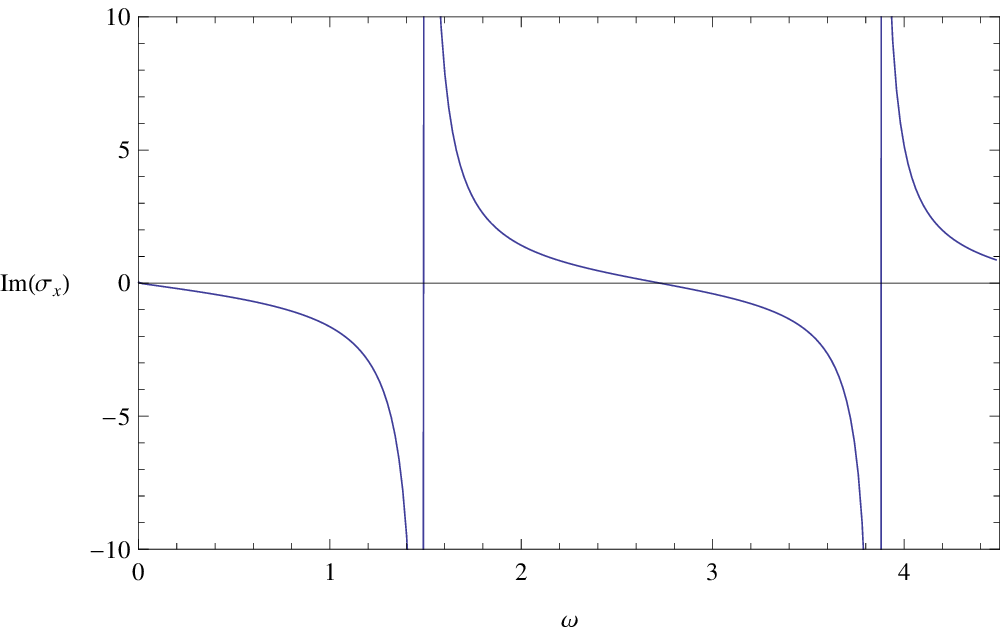}
\includegraphics[scale=0.7]{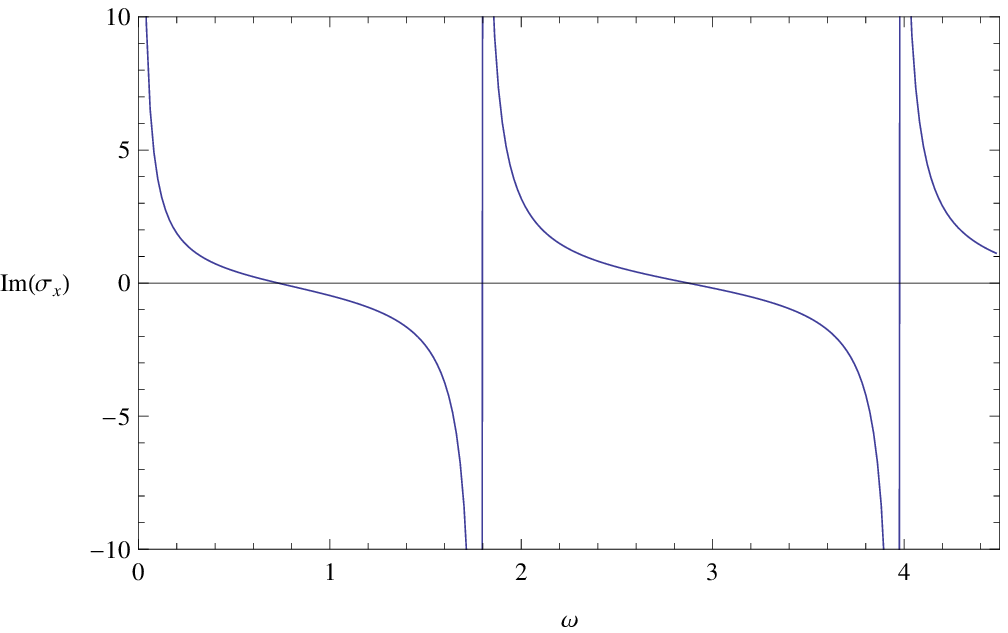}
\caption{\label{conducsoliton} The imaginary part
of the conductivity for the AdS soliton.}}
\end{figure}
The behavior of the conductivity in the AdS
soliton background presented here is similar to
the result when we did not consider the spatial
dependence of the vector potential in \cite{Q.Y. Pan-2}.

Furthermore we go on to study the  pair
susceptibility in the AdS soliton background.
Similarly, taking $\Psi=\psi(r)e^{-i\omega t}$,
the scalar perturbation equation reads
\begin{equation}\label{ps}
\psi''+(\frac{f'}{f}+\frac{2}{r})\psi'+[\frac{(A_t+\omega)^2}{f
r^2}-\frac{A_{\eta}^2}{f^2}-\frac{m^2}{f}]\psi=0.
\end{equation}
The asymptotic behavior near the boundary
$r\rightarrow \infty$ is still  $ \psi= \frac{\psi_1}{r}+ \frac{\psi_2}{r^{2}} +\cdots$. Beginning with (\ref{ps}), we can
find that the scalar field behaves as
 \begin{equation}\label{hh}
 \psi\simeq (r-r_0)^{| A_{\eta}|/3}[1+U (r-r_0)+V (r-r_0)^2+\cdots]
\end{equation}
at the soliton rip. It is interesting that in the
AdS soliton case, there is no imaginary part of
the dynamical pair susceptibility, only the real
part of pair susceptibility exists as shown in Fig.8
for the operator $\mathbb{O}_2$.  The vanish of
the imaginary part of the dynamical pair
susceptibility can be attributed to the fact that
both (\ref{ps}) and (\ref{hh}) are real.

Different from that in the AdS black hole, in the
AdS soliton case we observe that the real part of the
dynamical pair susceptibility is similar to the
description of the BCS pair instability \cite{F.
Marsiglio}. With the change of $1/\mu$, we also
observe the move of the curve with the horizontal
line up and down. This is similar to the effect
of the temperature in the description of the BCS
pair instability which is the key factor to
characterize the stability. However this is just
a phenomenological similarity at the first sight,
whether there is further deep connection still
needs careful study.

\begin{figure} \center{
\includegraphics[scale=0.7]{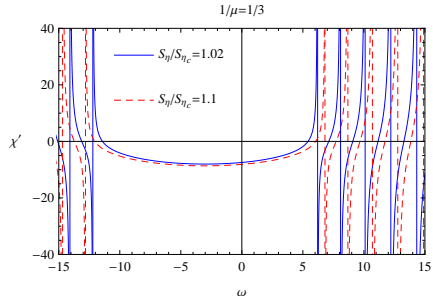}
\caption{\label{fig03}(color online) The real
part of the pair susceptibility for operators
$\mathbb{O}_2$.}}
\end{figure}

In conclusion, we have studied the gravity duals
of supercurrent solutions with general phase
structure to describe both the first and the
second order phase transitions at finite
temperature in strongly interacting systems. We
have argued that the conductivity and the pair
susceptibility, which are measurable quantities
in the condensed matter physics, can be possible
phenomenological indications to distinguish the
order of phase transitions. Besides the AdS black
hole background, we have also extended our
discussion to the AdS soliton configuration. We
have found that in the AdS soliton, the first
order phase transition cannot be brought by the
supercurrent. The conductivity behaves similar to
the case when there is only electric field $A_t$.
There is no imaginary part of the dynamical pair
susceptibility and the real part of the pair
susceptibility behaves similar to that disclosed
in the BCS pair instability \cite{F. Marsiglio}. Further
understanding on this phenomenological similarity
is called for.

\begin{acknowledgments}
This work has been supported partially by the
 NNSF of China and the Shanghai Science and
Technology Commission under the grant
11DZ2260700. We acknowledge Q.Y. Pan and Y. Peng
for helpful discussions.

\end{acknowledgments}


\begin{thebibliography}{99}

\bibitem{Maldacena}
J. M. Maldacena, Adv. Theor. Math. Phys. {\bf 2}, 231 (1998).

\bibitem{S.S.Gubser-1}
S. S. Gubser, I. R. Klebanov and A. M. Polyakov,
Adv. Theor. Math. Phys. Lett.B {\bf 428}, 105
(1998).

\bibitem{E.Witten}
E. Witten, Adv. Theor. Math. Phys. {\bf 2}, 253
(1998).
\bibitem{S.A. Hartnoll}
S.A. Hartnoll, Class. Quant. Grav. 26, 224002
(2009).

\bibitem{C.P. Herzog}
C.P. Herzog, J. Phys. A 42, 343001 (2009).

\bibitem{G.T. Horowitz-1}
G.T. Horowitz, arXiv:1002.1722 [hep-th].

\bibitem{G.T. Horowitz-2}
G.T. Horowitz and M.M. Roberts, Phys. Rev. D 78,
126008 (2008).

\bibitem{S.A. Hartnoll-1}
S.A. Hartnoll, C.P. Herzog, and G.T. Horowitz,
Phys. Rev. Lett. 101, 031601 (2008).

\bibitem{E. Nakano}
E. Nakano, W.Y. Wen, Phys. Rev. D 78, 046004
(2008).

\bibitem{I. Amado}
I. Amado, M. Kaminski, and K. Landsteiner, J.
High Energy Phys. 0905, 021 (2009).

\bibitem{G. Koutsoumbas}
G. Koutsoumbas, E. Papantonopoulos, and G.
Siopsis, J. High Energy Phys. 0907, 026 (2009).

\bibitem{O.C. Umeh}
O.C. Umeh, J. High Energy Phys. 0908, 062 (2009).

\bibitem{H.B. Zeng}
H.B. Zeng, Z.Y. Fan, and Z.Z. Ren, Phys. Rev. D
80, 066001 (2009).

\bibitem{J. Sonner-1}
J. Sonner, Phys. Rev. D 80, 084031 (2009).

\bibitem{S.S. Gubser-2}
S.S. Gubser, C.P. Herzog, S.S. Pufu, and T.
Tesileanu, Phys. Rev. Lett. 103, 141601 (2009).

\bibitem{J.P. Gauntlett}
J.P. Gauntlett, J. Sonner, and T. Wiseman, Phys.
Rev. Lett. 103, 151601 (2009).

\bibitem{R.G. Cai}
R.G. Cai and H.Q. Zhang, Phys. Rev. D 81, 066003
(2010).

\bibitem{J.L. Jing}
J.L. Jing and S.B. Chen, Phys. Lett. B 686, 68
(2010).

\bibitem{C.P. Herzog-1}
C.P. Herzog, Phys. Rev. D 81, 126009 (2010).

\bibitem{S.B. Chen}
S.B. Chen, L.C. Wang, C.K. Ding, and J.L. Jing,
Nucl. Phys. B 836, 222 (2010).

\bibitem{R.A. Konoplya}
R.A. Konoplya and A. Zhidenko, Phys. Lett. B 686,
199 (2010).

\bibitem{wjp}
J.P. Wu, Y. Cao, X. M. Kuang, W. J. Li, Phys.
Lett. B 697, 153-158 (2011).

\bibitem{G. Siopsis}
G. Siopsis and J. Therrien, J. High Energy Phys.
1005, 013 (2010).

\bibitem{K. Maeda}
K. Maeda, M. Natsuume, and T. Okamura, Phys. Rev.
D 79, 126004 (2009).

\bibitem{R. Gregory}
R. Gregory, S. Kanno, and J. Soda, J. High Energy
Phys. 0910, 010 (2009).

\bibitem{Q.Y. Pan-2}
Q.Y. Pan, B. Wang, E. Papantonopoulos, J.
Oliveira, and A.B. Pavan, Phys. Rev. D 81, 106007
(2010).


\bibitem{X.H. Ge}
X.H. Ge, B. Wang, S.F. Wu, and G.H. Yang, J. High
Energy Phys. 1008, 108 (2010).

\bibitem{X. He}
X. He, B. Wang, R.G. Cai, and C.Y. Lin, Phys.
Lett. B 688, 230 (2010).


\bibitem{R.G. Cai-1}
R.G. Cai, Z.X. Nie, B. Wang, and H.Q. Zhang,
arXiv:1005.1233 [gr-qc].

\bibitem{Kuang}
X.M. Kuang, W.J. Li and Y. Ling, J. High Energy
Phys. 1012, 069 (2010).

\bibitem{Amin Akhavana}
A. Akhavana, M. Alishahiha, Rev. D 83, 086003
(2011)


\bibitem{Y. Brihaye}
Y. Brihaye and B. Hartmann, Phys. Rev. D 83,
126008 (2010).


\bibitem{S.S. Gubser-3}
S.S. Gubser, Phys. Rev. D 78, 065034 (2008).



\bibitem{S.A. Hartnoll-2}
S. A. Hartnoll, C. P. Herzog, and G. T. Horowitz,
J. High Energy Phys. 0812 ,015 (2008).




\bibitem{S. Franco}
S. Franco, A.M. Garcia-Garcia, and D.
Rodriguez-Gomez, J. High Energy Phys. 1004, 092
(2010).


\bibitem{S. Franco-1}
S. Franco, A.M. Garcia-Garcia, and D.
Rodriguez-Gomez, Phys. Rev. D 81, 041901(R)
(2010).

\bibitem{ F.Aprile}
 F.Aprile and J.G. Russo, Phys. Rev. D 81, 026009 (2010).


\bibitem{Q.Y. Pan-1}
Q.Y. Pan and B. Wang, Phys. Lett. B 693, 159
(2010).

\bibitem{Yunqi Liu}
Y.Q. Liu, Q.Y. Pan and B. Wang, Phys. Lett. B 702
94-99 (2011).


\bibitem{Tatsuma}
T. Nishiokaa,S. Ryu and T. Takayanagi, J. High
Energy Phys. 1003, 131 (2010).

\bibitem{Gary T}
G. T.Horowitz, B. Way,  J. High Energy Phys.
1011, 011 (2010) .


\bibitem{Peng Yan}
P. Yan, Q.Y.Pan, B.Wang, Phys. Lett. B 699 (5)
(2011).

\bibitem{Bobev}
N. Bobev, A. Kundu, K. Pilch, N. P. Warner,  JHEP 1203 (2012) 064.

\bibitem{C. Herzog-2}
C. Herzog, P. Kovtun and D. Son, Phys. Rev. D 79,
066002 (2009).


\bibitem{P. Basu}
P. Basu, A. Mukherjee and H. H. Shieh, Phys. Rev.
D 79, 045010(2009).

\bibitem{J. Sonner}
J. Sonner and B. Withers, Phys. Rev. D 82,
026001(2010).

\bibitem{Daniel}
D. Arean, M. Bertolini, J. Evslin, T. Prochazka, JHEP 1007:060,2010.

\bibitem{P. Basu-1}
D. Arean, P. Basu and C. Krishnan, J. High Energy
Phys. 10, 006 (2010).

\bibitem{Binder}
K. Binder, Rep. Prog. Phys. 50 (1987) 783-859.

\bibitem{Horbach}
M. L. Horbach et al., Phys. Rev. B 46, 432 (1992); K.
Holczer et al., Phys. Rev. Lett. 67, 152 (1991).

\bibitem{K. Maeda}
K. Maeda, M. Natsuume, T. Okamura, Phys.Rev.D79,
126004 (2009).

\bibitem{YQ Liu-1}
Y.Q. Liu, Q.Y. Pan, B. Wang, R.G. Cai, Phys. Lett
B, 693(3) (2010).

\bibitem{She}
J.H. She, B. J. Overbosch, Y. W. Sun, Y. Liu, K.
Schalm, J. A. Mydosh and J. Zaanen, Phys. Rev. B
84, 144527 (2011).

\bibitem{Son}
D. T. Son and A. O. Starinets, J. High Energy
Phys.0209, 042 (2002).

\bibitem{KW}
I. R. Klebanov and E. Witten, Nucl. Phys. B 556, 89 (1999).

\bibitem{Montull1}
M. Montull, O. Pujolas, A. Salvio, P. J. Silva, Phys.Rev.Lett. 107 (2011) 181601

\bibitem{Montull2}
M. Montull, O. Pujolas, A. Salvio, P. J. Silva, JHEP 1204 (2012) 135

\bibitem{F. Marsiglio}
F. Marsiglio, K. S. D. Beach, and R. J. Gooding, [arXiv:1203.5122[cond-mat]].

\end{thebibliography}
\end{document}